\documentstyle[aps,epsf]{revtex}

\begin{document}
\title{CONDUCTIVITY FLUCTUATIONS IN POLYMER'S NETWORKS.}
\author{A.~ N.~ Samukhin${}^{1,2}$, V.~N.~Prigodin${}^{1,3}$, and L.~Jastrab\'\i k$%
{}^2$}
\address{$^1$A. F. Ioffe Physical \& Technical institute, 194021 St.Petersburg,\\
Russia.\\
$^2$Institute of Physics AS CR, Na Slovance 2, 180 40 Prague 8, Czech\\
Republic.\\
$^3$Max-Planck-Institut f\"ur Physik Komplexer Systeme, Au\ss enstelle\\
Stuttgart, Heisenbergstra\ss e 1, D-70569 Stuttgart, Germany.}
\maketitle

\begin{abstract}
Polymer's network is treated as an anisotropic fractal with fractional
dimensionality $D=1+\epsilon $ close to one. Percolation model on such a
fractal is studied. Using real space renormalization group approach of
Migdal and Kadanoff we find threshold value and all the critical exponents
in the percolation model to be strongly nonanalytic functions of $\epsilon $%
, e.g. the critical exponent of the conductivity was obtained to be $%
\epsilon ^{-2}\exp (-1-1/\epsilon )$. The main part of the finite size
conductivities distribution function at the threshold was found to be
universal if expressed in terms of the fluctuating variable which is
proportional to a large power of the conductivity, but with $\epsilon $%
-dependent low--conductivity cut-off. Its reduced central momenta are of the
order of $e^{-1/\epsilon }$ up to the very high order.
\end{abstract}

\draft

\bigskip

Disordered structures with fractional dimensionality arise in many various
physical applications \cite{bh92}. In particular, a number of papers are
devoted to the processes of diffusion on fractals \cite{nyo94,bmkse95,cr96}.
In our resent paper \cite{spj97} a new type of fractals is introduced: {\em %
nearly one-dimensional} strongly anisotropic ones. There we were dealing
with the problems of: i) percolation and ii) variable range hopping on them.
One motivation to study these problems were the experimental data on the
structure of some classes of oriented conducting polymers, and on their
unusual conducting and dielectric properties \cite{ts92,ar87,wjrme90,joea94}%
. The other is their conceptual significance, and, in particular, the
possibility to deal directly with the distribution functions (DF) of
strongly fluctuating random variables. We extend here an approach, developed
in \cite{spj97} in such a way, that we are able to obtain the DF of finite
sample conductivity on the percolation threshold.

In \cite{spj97} conducting polymers were modelled as fractal oriented
networks. Namely, we assume all chains to be oriented in one spatial
direction, and coupled transversely through various size perfectly
conducting islands. Fractional dimensionality was defined as follows: In an $%
L$-size cube chains form a set bundles, connected within this cube. If the
cross--section of maximal bundle scales as $L^\epsilon $ for large enough $L$%
, where $0\leq \epsilon \leq 2$, then we have $d^{*}=1+\epsilon $
--dimensional network. Obviously $\epsilon =0$ for purely one--dimensional
systems (sets of disconnected chains). The main feature of the fractals,
constructed from oriented 1d chains is their self-similarity: the system at
any scale looks like subdivided into bundles, which are in turn subdivided
into smaller ones, etc. Our hypothesis here is that oriented polymers
network structures are of this type (with $d^{*}=1+\epsilon $ close to 1, $%
\epsilon \ll 1$), at least in some wide enough length scales interval, e.g.
from the scale of polymer's fibrils (hundreds of nm) down to molecular
scales. Transmitting electronic micrographs (see e.g. \cite{ar87}) seems to
confirm this hypothesis.

A regular example of such a fractal is the hierarchical lattice \cite
{kd76,tm96}, constructed by the infinite repetition of two steps: i)
connection of $n$ bonds in sequence to form $n$-bond chain, and: ii)
assembling of $m$ $n$-chains in parallel to form $m$-bundle, which is
treated as a new bond on the next stage, etc. The fractal dimensionality of
such a system is: $d^{*}=\ln m/\ln n+1$. An exact real space renormalization
group may be written for hierarchical lattice, which becomes the
renormalization group of Migdal and Kadanoff (RGMK) \cite
{kd76,tm96,mg76,kdho75} if to set $n\rightarrow 1,\;m\rightarrow 1$ with the
value of $d^{*}$ fixed. Of course, the real polymers structures are not
regular ones, and the requirement of self-similarity here is to be treated
in statistical sense. Nevertheless, we shall use the RGMK scheme. The
additional argument for using this approach in nearly 1d case is that the
RGMK is exact in one dimension, therefore one may hope to obtain meaningful
results when the dimensionality is close to 1. This method was applied to
the percolation conductivity problem two decades ago by Scott Kirkpatrick 
\cite{k77}. He had found values of critical exponents of correlation length
and conductivity near threshold, using the RGMK equations for conductivity
momenta near threshold, truncated at the first moment. Though he had not
considered explicitly the case of dimensionality close to one, this method,
if appropriately applied, gives the right dependence of conductivity
exponent on $\epsilon $ up to pre-exponential factor.

The RGMK method may be formulated in a quite simple phenomenological
fashion. Suppose we have some random $d$-dimensional medium with the local
conductivity --- fluctuating random variable. Let us consider the $\lambda $%
-size cube within the medium. Its conductivity $\Sigma _\lambda $, and
resistivity $R_\lambda =1/\Sigma _\lambda $, are fluctuating random
variables with some DFs, defined (in the Laplace representation) as: 
\begin{equation}
P(\sigma ,\lambda )=\left\langle \exp \left( -\sigma \Sigma \left( \lambda
\right) \right) \right\rangle ;\quad Q(s,\lambda )=\left\langle \exp \left(
-sR\left( \lambda \right) \right) \right\rangle .  \label{def1}
\end{equation}
If we change the size of the cube, $\lambda \rightarrow \lambda ^{\prime
}=n\lambda $, we arrive at some new random variables $\Sigma (\lambda
^{\prime })$, $R(\lambda ^{\prime })$, the DFs are also changed, of course.
The cube's enhancement may be treated as a combination of $n$-times
expansions in one ``longitudal'' (arbitrarily chosen) spatial direction, and
in $d-1$ ``transverse'' ones. Further we shall transit to infinitesimal
transformation, so the order of operations is not important. For finite
length rescaling we {\em assume} that, changing the size $n$ times in the
longitudal direction, we arrive at $n$ resistivities in sequence, with the
resulting (specific) resistivity: 
\[
\widetilde{R}=\frac 1n\sum_1^nR_l\,. 
\]
where $R_l$ are {\em assumed} to be independent random variables. Therefore,
we have: 
\begin{equation}
\tilde Q(s)=\left\langle \exp \left( -s\tilde R\right) \right\rangle
=Q^n\left( s/n\right) \,,  \label{lg}
\end{equation}
and for infinitesimal transformation with $n=1+\delta \lambda /\lambda $,
the variation of the DF is: 
\begin{equation}
\delta _lQ(s,\lambda )=\left[ -s\frac{\partial Q(s,\lambda )}{\partial s}%
+Q(s,\lambda )\ln Q(s,\lambda )\right] \frac{\delta \lambda }\lambda \,,
\label{varlg}
\end{equation}
where $\delta _l$ means the variation due to longitudal rescaling. Quite
similarly, after the transverse rescaling, when the cross-section is
enlarged $m=n^{d-1}$ times, and the resulting conductivity is assumed to be
the arithmetic average of $m$ statistically independent ones, we have the
conductivities DF $P(\sigma )$ transforming in the same manner as in (\ref
{lg}), but with $n$ replaced by $m$. For infinitesimal transformation $%
m=1+\epsilon \,\delta \lambda /\lambda $, where $\epsilon =d-1$, and we have
for transverse rescaling: 
\begin{equation}
\delta _tP(\sigma ,\lambda )=\epsilon \left[ -\sigma \frac{\partial P(\sigma
,\lambda )}{\partial \sigma }+P(\sigma ,\lambda )\ln P(\sigma ,\lambda
)\right] \frac{\delta \lambda }\lambda \,.  \label{vartr}
\end{equation}

Using the integral identity: 
\[
e^{-x/\alpha }=1-\sqrt{x}\int_0^\infty \frac{dy}{\sqrt{y}}J_1\left( 2\sqrt{xy%
}\right) e^{-\alpha y}\,, 
\]
where $J_1$ is the Bessel's function, one can write the relation between
conductivities and resistivities DFs in the form of Hankel's transformation: 
\begin{equation}
Q(s,\lambda )=1-\sqrt{s}\int_0^\infty \frac{d\sigma }{\sqrt{\sigma }}%
J_1\left( 2\sqrt{s\sigma }\right) P\left( s,\lambda \right) \,.
\label{hankel}
\end{equation}
The reverse transformation is quite the same. Now we are able to write down
both transverse and longitudal variations in terms of either conductivities
or resistivities DF. Adding both variations for e.g. conductivity DF we
arrive at the following evolution equation upon size rescaling: 
\begin{eqnarray}
\ &&\lambda \frac{\partial P\left( \sigma ,\lambda \right) }{\partial
\lambda }=B\left( \left\{ P\right\} ,\sigma \right) =  \nonumber \\
&&\ \ \left( 1-\epsilon \right) \sigma \frac{\partial P(\sigma ,\lambda )}{%
\partial \sigma }+\epsilon P(\sigma ,\lambda )\ln P(\sigma ,\lambda )-\sqrt{s%
}\int_0^\infty \frac{ds}{\sqrt{s}}J_1\left( 2\sqrt{s\sigma }\right)
Q(s,\lambda )\ln Q(s,\lambda )\,.  \label{eveq}
\end{eqnarray}
This equation should be completed with Eq. (\ref{hankel}) to form a closed
set.

We may introduce the probabilities of $\lambda $-cube to be disconnected
(i.e. to have zero conductivity, or infinite resistivity) $c(\lambda )$
which, taking into account the definitions of DFs (\ref{def1}), may be
written as: 
\begin{equation}
c(\lambda )=P\left( +\infty ,\lambda \right) =1-Q\left( +0,\lambda \right)
\,.  \label{bb}
\end{equation}
Assuming in Eq. (\ref{eveq}) $\sigma =+\infty $, we have: 
\begin{equation}
\lambda \frac{dc}{d\lambda }=\beta _c\left( c\right) =\epsilon c\ln c-\left(
1-c\right) \ln \left( 1-c\right) \,.  \label{bbev}
\end{equation}
This equation has three fixed points: two stable ones, $c=0$ and $c=1$,
corresponding to connected and disconnected systems resp. in the
thermodynamic limit, and the unstable fixed point, $c=c_t$, $0<c_t<1$, 
\begin{equation}
\epsilon c_t\ln c_t=\left( 1-c_t\right) \ln \left( 1-c_t\right) \,,
\label{thrp}
\end{equation}
corresponding to the percolation threshold. The correlation length exponent $%
\nu $ is given by: 
\begin{equation}
\nu ^{-1}=\left. \frac{d\beta _c}{dc}\right| _{c=c_t}=\epsilon +1+\ln
c_t+\ln \left( 1-c_t\right) \,.  \label{nuexp}
\end{equation}
For nearly-1d systems, $\epsilon \ll 1$, we have: 
\begin{equation}
c_t=e^{-1/\epsilon },\;\;\;\nu =1/\epsilon \,.  \label{thrn1d}
\end{equation}

It is possible to rewrite Eq.(\ref{eveq}) using WKB-type approximation,
assuming: 
\begin{equation}
P\left( \sigma \right) =c+\left( 1-c\right) \exp \left[ -\phi (\sigma
)\right] \,,  \label{defphi}
\end{equation}
with $\phi \left( 0\right) =0$, and $\phi \left( \sigma \right) \rightarrow
\pm \infty $ as $\sigma \rightarrow \pm \infty $ rapidly enough (more
rapidly then $\pm \sqrt{\left| \sigma \right| }$, as we shall see later).
The important point also is to assume analyticity of $P\left( \sigma \right) 
$ and of $Q(s)$ at least within some finite width stripe along the real
axis. Using the relations: 
\[
J_1(z)=\frac{H_1^{(1)}(z)+H_1^{(2)}(z)}2\,,\quad H_1^{(1)}\left( ze^{i\pi
}\right) =-H_1^{(2)}(z)\,,\quad H_1^{(1)}(z)=-\frac{2i}{\pi z}\,\text{as }%
z\rightarrow 0\,, 
\]
where $H_1^{(1,2)}\,$ are the Hankel's function of first and second kind,
resp., and directing the cut of the function $H_1^{(1)}(2\sqrt{z})/\sqrt{z}$
in $z$ complex plane along positive real half-axis we may replace the
integrals with $J_1$-function along the positive real half-axis in Eqs. (\ref
{hankel}) and (\ref{eveq}) with the ones containing $H_1^{(1)}$, along the
following contour $C$: from $+\infty -i0$ to $\delta -i0$ along the bottom
shore of the cut, then from $\delta -i0$ to $\delta +i0$ along the almost
closed anticlockwise $\delta $-circle, and finally from $\delta +i0$ to $%
+\infty +i0$ along the top shore of the cut. Thus we have: 
\begin{equation}
Q(s)=-\left( 1-c\right) \frac{\sqrt{s}}2\int_C\frac{d\sigma }{\sqrt{\sigma }}%
H_1^{(1)}\left( 2\sqrt{s\sigma }\right) \exp \left( -\phi \left( \sigma
\right) \right)  \label{hank1}
\end{equation}
Assuming $\left| s\right| $ to be large enough, one may replace $H_1^{(1)}$
in the latter integral by its asymptotic expression: 
\[
H_1^{(1)}\left( 2\sqrt{s\sigma }\right) \simeq \pi ^{-1/2}\left( s\sigma
\right) ^{-1/4}\exp \left( -\frac{3i\pi }4+2i\sqrt{s\sigma }\right) \,, 
\]
and to treat this integral in the saddle point approximation. Afterwards,
the same procedure may be performed with the integral in Eq.(\ref{eveq}). As
a result, we have the evolution equation in the saddle point or ``WKB''
approximation to be: 
\begin{eqnarray}
\lambda \frac{\partial P}{\partial \lambda } &=&B_1\left( \left\{ P\right\}
,\sigma \right) =-\left( 1+\epsilon \right) \sigma P^{\prime }+\epsilon P\ln
P-\left( 1-c\right) \ln \left( 1-c\right) +\left( P-c\right) \ln \left(
P-c\right) -  \nonumber \\
&&\ \frac 12\left( P-c\right) \ln \left[ 1-2\sigma \frac{P^{\prime \prime }}{%
P^{\prime }}-2\sigma \frac{P^{\prime }}{P-c}\right] \,.  \label{evsp}
\end{eqnarray}

It seems that saddle point approximation is valid only at large enough
values of $\sigma $. The other approximation for the evolution equation is
possible if to set: $P\left( \sigma \right) =c+\left( 1-c\right) e^{-\sigma
}+\psi \left( \sigma \right) $, and to linearize Eq.(\ref{eveq}) with
respect to $\psi $. The remarkable fact is that after substitution of the
above expression into Eq.(\ref{evsp}), we arrive at the same linearized
equation for $\psi \left( \sigma ,\lambda \right) $. Thus, we have some
reason to look for an appropriate solution of Eq.(\ref{evsp}) in the whole
complex plane $\sigma $.

At the percolation threshold, $c=c_t$, one may to look for the solution of
the RG evolution equation in the form: $P(\sigma ,\lambda )=\bar P(\sigma
\lambda ^{-a})$, where $a=t/\nu $, $t$ and $\nu $ are critical exponents of
the conductivity, $\Sigma \propto \left( c_t-c\right) ^t$, and of
correlation length, $\xi \propto \left| c-c_t\right| ^{-\nu }$ . Then the
equation (\ref{evsp}) becomes an ordinary differential one of the second
order. It appears to be more convenient to use the function $\phi (x)=-\ln
\left[ \left( \bar P(x)-c_t\right) /(1-c_t)\right] $ instead of $\bar P(x)$.
Introducing $\phi _0=-\ln \left[ c_t/(1-c_t)\right] $, we have: 
\begin{equation}
\frac 12\ln \left[ 1+2x\frac{\phi ^{\prime \prime }}{\phi ^{\prime }}\right]
=\left( 1+\epsilon -a\right) x\phi ^{\prime }-\phi +\epsilon \left[ g\left(
\phi -\phi _0\right) -g\left( -\phi _0\right) \right] \,,  \label{steq}
\end{equation}
where $g(\phi )\equiv \left( e^\phi +1\right) \ln \left( 1+e^\phi \right) \,$%
. An equation for $\phi _0$ which follows from Eq. (\ref{thrp}) was used in
the derivation of Eq. (\ref{steq}). The latter may be easily solved after
the substitution: 
\begin{equation}
z\left( \phi \right) \equiv \exp \left[ -2\left( 1+\epsilon -a\right) x\phi
^{\prime }\right] \,,  \label{defz}
\end{equation}
with $\phi $ --- new independent variable. Requiring $z(\phi )\rightarrow 0$
as $\phi \rightarrow 0$ faster then $\exp \left[ -\left( 1+\epsilon
-a\right) \phi \right] $, we have: 
\begin{equation}
z\left( \phi \right) =\left( 1+\epsilon -a\right) e^{-\left( 1+\epsilon
-a\right) \phi }\int_\phi ^\infty dy\exp \left\{ -\left( 1+\epsilon
-a\right) y+2\epsilon \left[ g\left( y-\phi _0\right) -g\left( -\phi
_0\right) \right] \right\} \,.  \label{soln1}
\end{equation}
The normalization condition $\phi (0)=0$ implies $z(0)=1$, from which it
follows that: 
\begin{equation}
\left( 1+\epsilon -a\right) \int_0^\infty dy\exp \left\{ -\left( 1+\epsilon
-a\right) y+2\epsilon \left[ g\left( y-\phi _0\right) -g\left( -\phi
_0\right) \right] \right\} =1\,,  \label{eqa}
\end{equation}
which is an equation for $a$.

Comparing the values of $a$, obtained by the solution of Eq.(\ref{eqa}), and
by the numerical investigation of the evolution of the evolution equation (%
\ref{eveq}) \cite{spj97} one can see, that both methods give the same
results at any dimensionality. This, together with the considerations
presented above, prompts us to consider the saddle point solution as an
exact one. Of course, the RGMK method itself is an approximate one. In e.g.
three dimension we have from Eq. \ref{eqa}: $a\approx 1.891$. On the other
hand, the best possible at present numerical results \cite{nh97} give $%
a=2.25\pm 0.04$. So, the RGMK method may be not very bad even for 3-d
systems.

In case $\epsilon \ll 1$ our previous result \cite{spj97}: 
\begin{equation}
a=\frac{1+\epsilon }\epsilon \exp \left( -\frac{1+\epsilon }\epsilon \right)
\,,\;\;\;t=a/\epsilon \,,  \label{cexld}
\end{equation}
may be easily reproduced from Eq. (\ref{eqa}). In case of $\epsilon \gg 1$
one may obtain from Eq. (\ref{eqa}): 
\begin{equation}
a=\epsilon -\frac \epsilon 4e^{-\epsilon }\,.  \label{cexhd}
\end{equation}

Finally, the function $\phi (x)$ may be determined as a reverse of the
equation: 
\begin{equation}
Cx=\phi \exp \left[ -\int_0^\phi d\zeta \left( \frac{1+\epsilon -a}{z(\zeta )%
}-\frac 1\zeta \right) \right] \,.  \label{sol2}
\end{equation}
The arbitrary integration constant $C$ corresponds to arbitrary choice of
the unit of conductivity, or, alternatively, of the length scale at the
threshold point.

Thus the conductivities DF in the normal representation, $\Pi (\Sigma
(\Sigma ,\lambda )=c_t+(1-c_t)\bar \Pi \left[ \left( \lambda /\lambda
_0\right) ^a\Sigma \right] $, where the scaling function $\bar \Pi (y)$ may
be expressed as: 
\begin{equation}
\bar \Pi (y)=\int_{-i\infty }^{+i\infty }\frac{dx}{2\pi i}\exp \left[
xy-\phi (x)\right] =\frac 1y\int_{-i\infty }^{+i\infty }\frac{d\phi }{2\pi i}%
\exp \left[ -\phi +yx(\phi )\right] \,,  \label{scf}
\end{equation}
the last equality was obtained through the integration by parts. However,
there is some trouble when evaluating integral in Eq. (\ref{scf}). Namely:
the function $x(\phi )$ is singular at $\phi =\tilde \phi _n=\phi _0+i\pi
(2n+1)$, $n$- integer. The origin of these singularities is in the procedure
of analytic continuation made when the RGMK approach was formulated. It can
be illustrated as follows: Let us assume the initial distribution of
conductivities to be: $P_0(\sigma )=c+(1-c)e^{-\sigma }\,.$After putting $m$
identically distributed conductivities in parallel, the Laplace of the DF of
their sum, $P_1(\sigma )=\left[ c+(1-c)e^{-\sigma }\right] ^m\,$, has $m$-th
order zeroes at $\tilde \sigma _n=\sigma _0+i\pi (2n+1)$, $n$-integer, $%
\sigma _0=\ln [(1-c)/c]$, which turns into singularities after analytic
continuation to noninteger $m$. This points us, that it is the procedure of
the transition from integer rescaling factor transformation (which is exact
for a hierarchical structure) to the infinitesimal one (which no explicit
structure corresponds to) that is the reason of these singularities. So,
these singularities are to be treated as artifficial ones, and should be
avoided during the integration in Eq. (\ref{scf}). In general, this
restricts our knowledge of the DF $\bar \Pi \left( y\right) $ with its low-
and large-conductivities asymptotic behavior.

At large scaled conductivities $y$, shifting integration contour in Eq. (\ref
{scf}) to the region $\Re \phi >\phi _0$, one has the following asymptotic
expression for the DF: 
\begin{equation}
\bar \Pi (y)=\frac{D_2}{y_2}\left( \frac y{y_2}\right) ^{\frac{1+\epsilon }{%
2a}-1}\exp \left[ -\left( \frac y{y_2}\right) ^{\frac{1+\epsilon }a}\right]
\,,  \label{leftscf}
\end{equation}
where: 
\begin{eqnarray}
D_2 &=&\frac{[2\pi (1+\epsilon )(1+\epsilon -a)]^{1/2}}{(1-c_t)a}\left( 
\frac{1+\epsilon +a}{1+\epsilon -a}\right) ^{\frac 1{2(1+\epsilon )}}\,,\ \
y_2=\frac{1+\epsilon }{1+\epsilon -a}\left( \frac{1+\epsilon -a}a\right) ^{%
\frac a{1+\epsilon }}e^{A_2}\,,  \nonumber \\
A_2 &=&2(1+\epsilon -a)\int_{-\infty }^0d\zeta \ln (-\zeta )\frac d{d\zeta }%
\frac \zeta {\ln z(\zeta )}  \label{leftpar}
\end{eqnarray}
Shifting the integration contour in Eq.(\ref{scf}) to the region $\Re \phi
>\phi _0$, we arrive at the following expression for the $\bar \Pi (y)$ DF
at small $y$-region: 
\begin{equation}
\bar \Pi (y)=\frac{D_1}{y_1}\left( \frac y{y_1}\right) ^{\frac 1{2(\epsilon
-a)}+1}\exp \left[ -\left( \frac y{y_1}\right) ^{\frac 1{\epsilon -a}%
}\right] \,,  \label{rightscf}
\end{equation}
with: 
\begin{eqnarray}
D_1 &=&\frac{[2\pi (1+\epsilon -a)]^{1/2}}{\epsilon -a}e^{-\epsilon
}c_t^{-\epsilon /(1-c_t)}\,,\ \ y_1=\frac{e^{-A_1}}{1+\epsilon -a}\left( 
\frac{\epsilon -a}{1+\epsilon -a}\right) ^{\epsilon -a}  \nonumber \\
A_1 &=&2(1+\epsilon -a)\int_0^\infty d\zeta \ln \zeta \frac d{d\zeta }\frac %
\zeta {\ln z(\zeta )}\,,  \label{rightpar}
\end{eqnarray}

More detailed results are available in the limiting case $\epsilon \ll 1$.
Here we have at $\Re \phi <\phi _0\simeq 1/\epsilon $, the following
expression for $x(\phi )$, truncated at the first order of $c_t$ and of $a$: 
\begin{equation}
\ln x(\phi )=\ln \phi +c_t\frac{e^\phi -1-\phi }\phi +\frac a{(1+\epsilon )^2%
}\int_0^\phi \frac{d\zeta }{\zeta ^2}\left[ e^{(1+\epsilon )\zeta
}-1-(1+\epsilon )\zeta \right] \,.  \label{leftx1d}
\end{equation}
Evaluating Taylor series of $\phi \left( x\right) $ at $x=0$, we obtain
central momenta of the conductivity to be of the order of $a$: 
\begin{eqnarray}
\frac{\left\langle \Sigma -\left\langle \Sigma \right\rangle \right\rangle ^2%
}{\left\langle \Sigma \right\rangle ^2}=a+c_t\,,\ \ \ \frac{\left\langle
\Sigma -\left\langle \Sigma \right\rangle \right\rangle ^3}{\left\langle
\Sigma \right\rangle ^3}=-\frac 12(1+\epsilon )a+c_t\,,\ \ldots \ 
\label{mom1d}
\end{eqnarray}

\begin{figure}[tbp]
\epsfxsize=12cm
\epsfbox{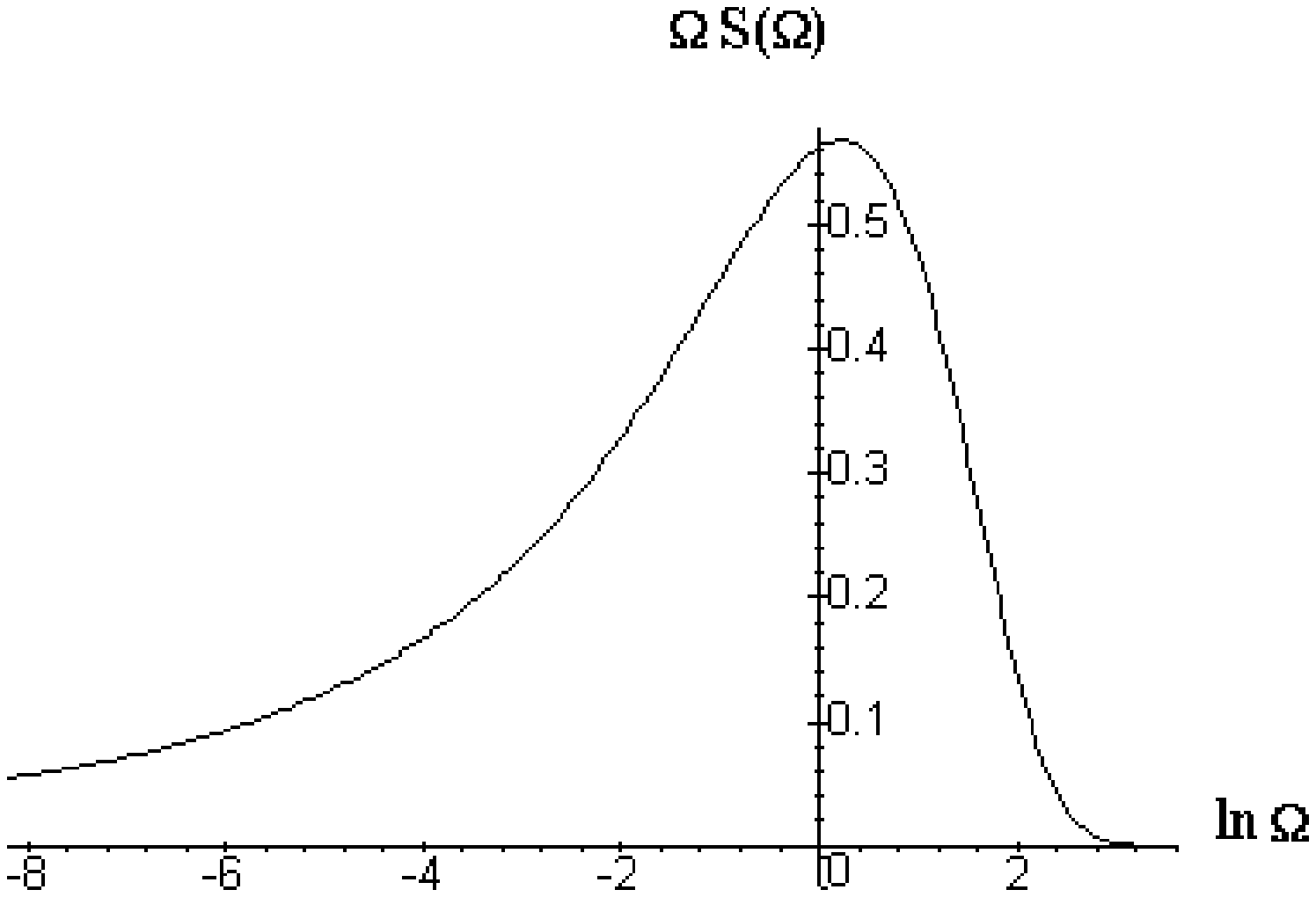}
\caption{Limiting form of the distribution function for the rescaled
logariphm of the conductivity.}
\label{univ}
\end{figure}

On the other hand, using in Eq.(\ref{scf}) the asymptotics of $x(\phi )$ at $%
\Re \phi <0$, $\left| \phi \right| \gg 1$, which may be justified at
sufficiently large $y$, we have, after the proper change of the integration
variable: 
\begin{equation}
y\bar \Pi (y)=e^{\phi _2}\frac{1+\epsilon }a\Omega S(\Omega )\,,
\label{1daslarge}
\end{equation}
where $\phi _2=a/(1+\epsilon )^2-c_t$, and the new fluctuating variable was
introduced: 
\begin{equation}
\Omega =e^{G_1}\frac a{1+\epsilon }y^{\frac{1+\epsilon }a}\,,\ \ \
G_1=1-\gamma -\ln (1+\epsilon )-\frac{(1+\epsilon )c_t}a\simeq 1-\gamma \,,
\label{Omegay}
\end{equation}
$\gamma $ is the Euler's constant, and $S(\Omega )$ is given by: 
\begin{equation}
S(\Omega )=\int_{-i\infty +\Delta }^{i\infty +\Delta }\frac{du}{2\pi i}%
u^{\Omega u}=\int_{-\pi }^\pi \frac{d\theta }{2\pi }V(\theta )e^{-\Omega
V(\theta )}\,,\;\;V(\theta )=\frac \theta {\sin \theta }e^{-\theta \cot
\theta }\,.  \label{sfunc}
\end{equation}
The latter expression for $S(\Omega )$ in (\ref{sfunc}) was obtained
choosing the integration contour along the line $\Im u=0$ in the $u$ complex
plane. Asymptotical expressions for $S(\Omega )$ may be easily obtained by
the saddle-point method: 
\begin{equation}
S(\Omega )\approx \left\{ 
\begin{array}{l}
\frac{\exp (-e^{-1}\Omega )}{\sqrt{2\pi e\Omega }},\text{~as }\Omega \gg 1\,;
\\ 
\frac{\sqrt{2\pi }}{e\Omega }\left[ \frac{\ln \ln (e/\Omega )}{\ln (e/\Omega
)}\right] ^2,\text{~as }\Omega \ll 1.
\end{array}
\right.  \label{sfuncas}
\end{equation}
Figure \ref{univ} shows $\Omega S\left( \Omega \right) $ as a function of $%
\ln \Omega $.

The asymptotics of $\bar \Pi (y)$ at small $y$ is defined by the expressions
(\ref{rightscf}, \ref{rightpar}). After some simple calculations we have: 
\begin{equation}
y\bar \Pi (y)=\frac{y^{-\frac 1{2\epsilon }}}{\sqrt{2\pi }\epsilon }\exp
\left( \frac 12-e^{-1}y^{-1/\epsilon }\right) \,.  \label{left1das}
\end{equation}
The two expressions, (\ref{1daslarge}) and (\ref{left1das}), can be sewed
together using the expression in the intermediate region, where the function
in the integral in Eq. (\ref{scf}) can be expanded up to the first order in $%
a$, $c_t$. This yields in the region $0<y<1$: 
\begin{equation}
\bar \Pi (y)=\frac a{1+\epsilon }\frac 1{(1-y)^2}\,.  \label{1dinm}
\end{equation}

To establish regions of validity for three expression of the DF, let us set $%
y=1-\Delta $, $\Delta \ll 1$. Comparing Eq.(\ref{1dinm}) with Eqs.(\ref
{1daslarge}) and (\ref{left1das}), one can find, that Eq.(\ref{1daslarge})
is valid if $\Delta <\Delta _1\sim a/\epsilon \ln \left( 1/\epsilon \right) $%
, and Eq.(\ref{left1das}) --- if $\Delta >\Delta _2\sim \epsilon \ln \left(
e/\epsilon \right) $. Let us note, that in spite of the low-conductivity
cut-off for the universal distribution (\ref{1daslarge}) is very close to
the mean value in terms of the conductivity itself, $\Delta _1\ll 1$, which
ensures central momenta of the conductivity to be of the order of $a$, this
cut-off is small in terms of the universally fluctuating variable $\Omega $, 
$\Omega _1\sim a^p$, $p\approx 1+1/\ln \left( 1/\epsilon \right) $.

The distribution function $S\left( \Omega \right) $ arise naturally in 1d
chain of random resistors, if to require scaling form of the distribution
function of $\lambda $-length chain specific resistivities: $\Upsilon \left(
R,\lambda \right) =\bar \Upsilon (R\lambda ^{-a})$, or $Q\left( s,\lambda
\right) =\bar Q\left( s\lambda ^a\right) $ in the Laplace representation.
Then from $Q\left( s,n\lambda \right) =Q^n\left( s/n,\lambda \right) $ one
immediately has: $\bar Q\left( x\right) =\exp \left[ -Cx^{1/\left(
1+a\right) }\right] $. Evaluating its inverse Laplace $\Upsilon \left(
r\right) $, and assuming $a\ll 1$, which is true in 1d case, we have after
the proper rescaling of the integration variable: 
\begin{equation}
r\Upsilon \left( r\right) =\frac 1a\Omega S\left( \Omega \right) ,\;\;\Omega
=ar^{-1/a},  \label{scf1d}
\end{equation}
which is essentially the same formula as Eqs. (\ref{1daslarge},\ref{Omegay}).

It should be noted that, due to the nature of approximations used in the
derivation of the RGMK equations, it is the distribution function at large
resistivities (or low conductivities), which is most suspected to be
inadeqately reproduced. Indeed, in reality some of the percolative paths at
the threshold infinite cluster inevitably have return parts, where the
diffusive particle moves in opposite direction. Such a paths are not taken
into account in the RGMK scheme, which can be especially clearly
demonstrated using hierarchical structures approach in the derivation \cite
{spj97}. This leads us e.g. to the overestimation of the conductivity value,
resulting in lower value of the exponent $a=t/\nu $, which we found to be $%
1.890$ vs. $2.25\pm 0.04$ obtained through numerical simulation \cite{nh97}
in 3d, and $0.818$ vs. $0.98$ in 2d. Still the main part of conductivities
distribution function, described by Eq. (\ref{1daslarge}), seems to be valid
in nearly-1d case.

In conclusion let us note that the method suggested enables one to deal not
only with distribution functions of conductivities in the percolative
systems --- it may also be applied to treat fluctuations of random variables
in other disordered systems, e.g. ones described by random coupling Ising
and Potts models.

The authors thanks V.V. Bryksin, Yu.A. Firsov, S.N. Dorogovtsev, B.N.
Shalaev and W. Wonneberger for stimulating discussions. This work was
partially supported by a Russian National Grant No RFFI 96-02-16848-a.

\begin{references}
\bibitem{bh92}  {\em Fractals and Disordered Systems}, Ed. by A. Bunde, S.
Havlin (Springer Verlag, Berlin-Heidelberg-New-York, 1992).

\bibitem{nyo94}  T.~Nakayama, K.~Yakubo and R.~L.~Orbach, Rev. Mod. Phys., 
{\bf 66}, 381, (1994).

\bibitem{bmkse95}  M. Ben-Chorin, F. Moller, F. Koch, W. Schirmacher and M.
Eberhard, Phys. Rev. B, {\bf 51}, 2199 (1995).

\bibitem{cr96}  D. Cassi, S. Regina, Phys. Rev. Lett., {\bf 76}, 2914, 1996.

\bibitem{spj97}  A.N. Samukhin, V.N. Prigodin, L. Jastrab\'{\i }k, Phys.
Rev. Lett, {\bf 78}, 326, 1997.

\bibitem{ts92}  J.~Tsukamoto, Adv. Phys., {\bf 41}, 509 (1992).

\bibitem{ar87}  K. Araya, T. Micoh, T. Narahara, K. Akagi and H. Shirakawa,
Synth. Metals, {\bf 17}, 247 (1987).

\bibitem{wjrme90}  Z. H. Wang, H.\ H.\ S. Javadi, A. Ray, A. J. MacDiarmid
and A. J. Epstein, Phys. Rev. B, {\bf 42}, 5411 (1990).

\bibitem{joea94}  J. Joo, Z. Oblakowski, G. Du, J. P. Pouget, E. J. Oh, J.
M. Weisinger, Y. Min, A. G. MacDiarmid and A. J. Epstein: Phys. Rev. B, {\bf %
49}, 2977 (1994).

\bibitem{kd76}  L.~P.~Kadanoff, Ann. Phys. (N.Y.), {\bf 100}, 359 (1976).

\bibitem{tm96}  C. Tsallis and A. C. N. de Magalh\~{a}es, Phys. Repts., {\bf %
268}, 305 (1996).

\bibitem{mg76}  A.~A~.Migdal, Sov. Phys. JETP, {\bf 42}; 413, 743 (1976).

\bibitem{kdho75}  L.~P.~Kadanoff and A.~Houghton, Phys. Rev. B, {\bf 11},
377 (1975).

\bibitem{k77}  S.~Kirkpatrick, Phys. Rev. B, {\bf 15}, 1533 (1977).

\bibitem{nh97}  J-P. Normand and H. J. Herrmann, Int. J. Mod. Phys. C, to be
published.
\end{references}
\end{document}